\documentclass[a4paper]{article}
\usepackage{ISCSLP2026}
\usepackage{ifthen}
\usepackage{enumitem,multirow,url}
\newboolean{blind}
\setboolean{blind}{false} 
\title{Neural Music Enhancement with Dual Time-Frequency Spectral Representations for Prediction and Discrimination}
\name{
	\ifthenelse{\boolean{blind}}{Anonymous to ISCSLP}
	{Fei Liu, Yang Ai$^*$\thanks{$^*$Corresponding author. This work was supported by the National Key Research and Development Program Project 2024YFE0217200, and the National Natural Science Foundation of China under Grant 62301521.}, Zhen-Hua Ling}
}
\address{
  \ifthenelse{\boolean{blind}}{Anonymous to ISCSLP}
  {
  	National Engineering Research Center of Speech and Language Information Processing,\\ 
University of Science and Technology of China, Hefei, China
  }
}

\email{
	\ifthenelse{\boolean{blind}}{Anonymous to ISCSLP}
	{fliu215@mail.ustc.edu.cn, yangai@ustc.edu.cn, zhling@ustc.edu.cn}
}

\begin{document}

\maketitle
\begin{abstract}
Non-professional music recordings shared online often suffer from background noise and reverberation, degrading perceived quality and limiting reuse. This paper proposes DSME, a music enhancement model based on dual time-frequency spectral representations. Within a generative adversarial framework, DSME uses short-time Fourier transform (STFT) spectra for generation and constant-Q transform (CQT) spectra for discrimination. Leveraging STFT's fixed window, invertibility, and predictability, the generator estimates clean amplitude-phase spectra from degraded inputs and reconstructs waveforms via inverse STFT. Exploiting CQT's log-frequency, variable-window structure aligned with musical octaves, we design an octave-segmented CQT discriminator. We also introduce a chroma-spectrum loss to emphasize pitch and harmonic consistency. Experiments show DSME outperforms baselines in objective and subjective tests, validating the effectiveness of the dual-spectrum approach.
\end{abstract}

\vspace{-1mm}
\section{Introduction}
\vspace{-1mm}

With the rapid growth of social media and video-sharing platforms, the internet is increasingly filled with music recorded in non-professional environments. Such user-generated content is often captured with consumer-grade devices in uncontrolled settings, resulting in issues like background noise and reverberation. Although these recordings are generally intelligible, the music quality is significantly reduced, impacting the listening experience and motivating music enhancement for applications such as tagging, recommendations, and transcription \cite{casey2008content}. Compared to speech enhancement, music enhancement faces additional challenges due to polyphonic sources, greater timbral diversity, and stricter perceptual fidelity requirements \cite{schaffer2022music,tachibana2013singing}.

Most music enhancement research has extended speech-oriented techniques. Traditional signal processing approaches and early neural methods have been applied, yet often fall short in complex music scenarios \cite{boll2003suppression,chen2006new,le2012consistent}. More recently, deep learning–based music denoising models have been proposed, including U-Net–style encoder–decoder networks and attention-augmented architectures such as MusicECAN \cite{cheng2024musicecan,moliner2022behm,moliner2022two,shao2024music}. Kandpal \MakeLowercase{\textit{et al.}} \cite{kandpal2022music} used a Pix2Pix-based network \cite{isola2017image} to improve mel-spectrograms and a DiffWave vocoder \cite{kong2020diffwave} for waveform synthesis, achieving strong performance. Building on this line, Chae \MakeLowercase{\textit{et al.}} \cite{chae2023exploiting} leveraged the encoder–decoder structure of CMGAN \cite{abdulatif2024cmgan} and introduced a time-frequency Conformer module to better capture temporal-spectral dependencies.

\begin{figure*}[t]
    \centering
    \includegraphics[width=0.85\linewidth]{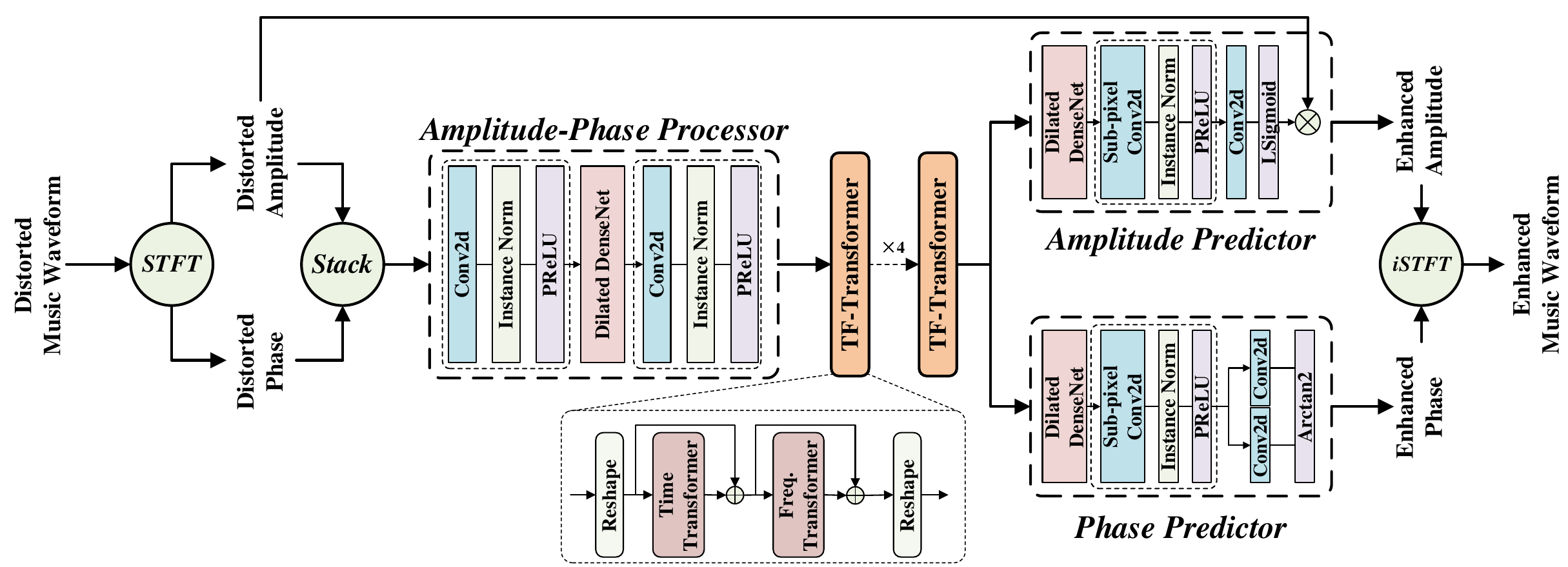}
    \caption{Overview of the amplitude-phase-prediction-based generator of DSME.}
    \label{fig:g}
\end{figure*}

However, existing methods still have notable limitations: early U-Net variants are often restricted to additive noise, and recent approaches largely model mel-spectrograms or complex spectra without explicitly and accurately predicting phase, leaving room for improvement for music with rich harmonics, complex rhythms, and an expansive dynamic range. To address these challenges, we propose DSME, a neural model that leverages dual time-frequency spectral representations within a generative adversarial network (GAN) framework \cite{goodfellow2014generative}. Specifically, DSME uses the short-time Fourier transform (STFT) spectrum for explicit amplitude–phase prediction, benefiting from its fixed-window, invertibility, and predictability, while the discriminator operates on the constant-Q transform (CQT) spectrum whose log-frequency scale and variable window better capture octave-related harmonic structures via octave-segmented analysis. We further introduce a chroma-spectrum-based loss to preserve pitch and harmonic consistency. Experimental results show that DSME outperforms baseline models in objective and subjective evaluations across noise, reverberation, and mixed-distortion scenarios, confirming the effectiveness of the proposed dual-spectrum approach.

\vspace{-1mm}
\section{Proposed Method}
\vspace{-1mm}
The proposed DSME aims to restore clean music from degraded music within a GAN framework, where the generator predicts clean amplitude–phase spectra, while the discriminator operates on the CQT spectrum for discrimination. 

\vspace{-2mm}
\subsection{Amplitude-Phase-Prediction-based Generator}
\vspace{-2mm}

STFT offers linearly spaced bins, a fixed window, and perfect invertibility, providing a structured time-frequency representation well suited to neural prediction. Since phase is crucial in music, DSME’s generator uses an explicit amplitude-phase STFT representation.
The overall structure of the generator is shown in Fig. \ref{fig:g}. 
Inspired by \cite{lu2023mp,lu2025explicit}, the distorted music waveform $\bm{y}\in \mathbb{R}^{L}$ is first transformed into an amplitude spectrum $\bm{A}^{(y)}\in \mathbb{R}^{N \times K}$ and a phase spectrum $\bm{P}^{(y)}\in \mathbb{R}^{N \times K}$, which are stacked as $\bm{S}^{(y)}\in \mathbb{R}^{N \times K \times 2}$, where $L$ is the waveform length, $N$ and $K$ are the number of frames and frequency bins for spectra, respectively.
Subsequently, the feature $\bm{S}$ is processed by an amplitude-phase processor and 4 time-frequency (TF) Transformers, and fed into separate amplitude and phase predictors to estimate the clean amplitude spectrum $\bm{\hat{A}}^{(x)}\in \mathbb{R}^{N \times K}$ and clean phase spectrum $\bm{\hat{P}}^{(x)}\in \mathbb{R}^{N \times K}$.
Finally, the enhanced music waveform $\bm{\hat{x}}\in \mathbb{R}^{L}$ is reconstructed through an inverse STFT (iSTFT).
The details of each module are as follows.

\begin{itemize}[leftmargin=*]
\item {}{\textbf{Amplitude-Phase Processor}:} 
To fuse the degraded amplitude and phase spectra, we employ a dilated DenseNet \cite{iandola2014densenet} with two surrounding convolutional blocks (Conv2d + instance normalization + PReLU). The DenseNet stacks dilated convolution layers with dense concatenation of all previous feature maps, enlarging the time-axis receptive field and promoting feature reuse.
\item {}{\textbf{TF-Transformers}:}
The TF-Transformers further refine the features for subsequent prediction. Each TF-Transformer block contains a time-Transformer and a frequency-Transformer with identical structure and residual connections. We adopt a GRU-driven Transformer without positional encoding, consisting of multi-head self-attention and a GRU-based position-wise feed-forward network, both equipped with layer normalization and residual connections.
\item {}{\textbf{Amplitude \& Phase Predictor}:}
The two predictors explicitly output clean amplitude and phase spectra, sharing the same front-end: a dilated DenseNet followed by a deconvolutional block (sub-pixel Conv2d + instance normalization + PReLU). The amplitude head uses a mask estimation module (Conv2d + LSigmoid) and applies the mask to the distorted amplitude via element-wise multiplication to obtain $\bm{\hat{A}}^{(x)}$.
The phase head adopts a parallel estimation architecture \cite{ai2023neural}, predicting pseudo-real $\hat{\bm{P}}_r^{(x)} \in \mathbb{R}^{N \times K}$ and pseudo-imaginary  $\hat{\bm{P}}_i^{(x)} \in \mathbb{R}^{N \times K}$ with two Conv2d layers, then converting them to phase via the two-parameter arctangent function. 
\end{itemize}

\begin{figure*}[t]
    \centering
    \includegraphics[width=0.78\linewidth]{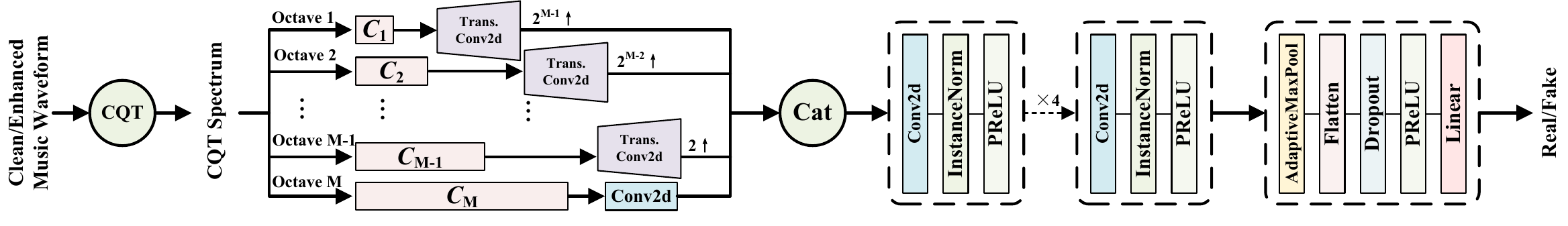}
    \caption{Overview of the octave-segmented CQT spectrum discriminator of DSME.}
    \label{fig:d}
\end{figure*}

\vspace{-2mm}
\subsection{Octave-Segmented CQT Spectrum Discriminator}
\vspace{-2mm}

Music notes comprise 12 semitones per octave with adjacent frequencies scaled by $2^{1/12}$.CQT bins align naturally with this semitone spacing, making CQT well suited for music representation. Since CQT is non-invertible, we do not use it for prediction; instead, we use it in training to build the octave-segmented CQT discriminator, as illustrated in Fig. \ref{fig:d}.

\begin{itemize}[leftmargin=*]
\item {}{\textbf{Octave-Segmented CQT Calculation}:} 
The discriminator first computes the CQT spectrum of either the clean music waveform $\bm{x}\in\mathbb R^L$ or the enhanced waveform $\hat{\bm{x}}\in\mathbb R^L$. 
An octave-segmented CQT spectrum is constructed to render the input more amenable to processing by the discriminator network. 
Consider a music waveform $\bm{z}\in\mathbb R^L$ ($\bm{z}=\bm{x}$ or $\hat{\bm{x}}$) sampled at $f_s$ Hz, with a lowest CQT center frequency of $f_{min}$ Hz. 
Suppose the CQT spans $M$ octaves with $B$ frequency bins per octave. 
The center frequency of the $k$-th frequency bin is then given by: 
\begin{equation}
    f_k=f_{min} \cdot 2^{\frac{k}{B}},
\end{equation}
where $k=0,1,\dots,BM-1$. 
Therefore, the ratio between each center frequency and its bandwidth is a constant $Q$, i.e.,
\begin{equation}
    Q=\frac{f_k}{\Delta f_k}=\frac{1}{2^{\frac{1}{B}}-1}.
\end{equation}
The CQT window length for the $k$-th frequency bin is given by:
\begin{equation}
    T_k^{(wl)}=\left\lceil \frac{Q\cdot f_s}{f_k} \right\rceil,
\end{equation}
and the hop size is kept constant within each octave and is defined as
\begin{equation}
    T_k^{(fs)}=2^{M+1-\lfloor \frac{k}{B}\rfloor}.
\end{equation}
Accordingly, for the $m$-th octave, at the $n$-th frame and the $k$-th frequency bin, the CQT spectral coefficient is given by
\begin{equation}
\bm{C}_m[n,k] = \sum_{t=0}^{T_k^{(wl)}-1} \bm{z}[n\cdot T_k^{(fs)}+t] \cdot \bm{w}_k[t]\cdot e^{-j2\pi t\cdot\frac{f_k}{f_s}},
\end{equation}
where $n=0,1,\dots,\left\lceil {L}/{T_k^{(fs)}} \right\rceil-1$, $k=0,1,\dots,B-1$ and $m=1,2,\dots,M$. 
$\bm{w}_k$ is the Hanning window with length of $T_k^{(wl)}$. 
Consequently, the resulting CQT spectrum is $\bm{C}_m\in\mathbb C^{\left\lceil {L}/{T_k^{(fs)}} \right\rceil\times B}, m=1,2,\dots,M$. 

\item {}{\textbf{CQT Spectrum Discrimination}:} 
Next, we feed the CQT spectrum $\bm{C}_1,\dots,\bm{C}_M$ into the discriminator network to determine whether it is real or generated. 
Specifically, the complex-valued CQT spectrum is first converted into a real-valued representation by concatenating its real and imaginary parts, i.e., $\bm{C}^{(RI)}_m=\text{Concat}[Re(\bm{C}_m),Im(\bm{C}_m)]\in \mathbb R^{\left\lceil {L}/{T_k^{(fs)}} \right\rceil\times 2B}, m=1,2,\dots,M$. 
Due to the different temporal resolutions of the CQT spectra across octaves, we upsample $\bm{C}^{(RI)}_m, m=1,\dots,M-1$ by a factor of $2^{M-m}$ using separate 2D transposed convolutions so that they match the temporal resolution of $\bm{C}^{(RI)}_M$, while $\bm{C}^{(RI)}_M$ is processed only by a standard 2D convolution. 
The processed features with matched temporal resolution are then concatenated and fed into 4 convolutional blocks, each consisting of a 2D convolution, an instance normalization layer, and a PReLU activation. 
The resulting feature map is passed through an adaptive max-pooling layer and flattened, then fed into a lightweight fully connected classifier with spectral normalization, PReLU nonlinearity, and dropout for further refinement. 
Finally, a learnable linear layer produces a scalar discriminator score.
\end{itemize}

\vspace{-3mm}
\subsection{Training Strategy}
\vspace{-2mm}

The training of DSME adopts a generative adversarial strategy and incorporates additional spectral-level losses.
\vspace{-2mm}
\subsubsection{Generative Adversarial Loss}
\vspace{-2mm}
We adopt a hinge-based generative adversarial loss, whose forms for the generator and the discriminator are respectively given as
$\mathcal L_{G}$ and $\mathcal L_{D}$.
In addition, we apply a feature matching loss $\mathcal L_{FM}$ to stabilize adversarial training and promote alignment between clean and enhanced waveform representations.

\vspace{-2mm}
\subsubsection{Spectral-Level Loss}
\vspace{-2mm}
In addition to the generative adversarial loss, we also introduce several STFT-domain spectral losses, including:

\begin{itemize}[leftmargin=*]
\item {}{\textbf{Amplitude Loss}:}
The amplitude loss $\mathcal L_{A}$ is defined as the mean square error (MSE) between the amplitude spectrum of the clean music waveform and that of the enhanced output.

\item {}{\textbf{Chroma Loss}:}
We further introduce a music-oriented chroma loss to better exploit musical structure and improve the perceptual quality of the enhanced music. 
The chroma spectrum is derived from the amplitude spectrum by compressing it into 12 pitch classes corresponding to the chromatic scale. 
Specifically, let $\bm{H}\in\mathbb R^{K\times 12}$ denote the chroma filterbank (i.e., a weight matrix grouped by pitch class). 
The chroma spectra of the clean music and enhanced one are then defined as
\begin{equation}
    \bm{R}=\bm{A}^\gamma\bm{H}, \quad\hat{\bm{R}}=\hat{\bm{A}}^\gamma\bm{H},
\end{equation}
where $\bm{R},\hat{\bm{R}}\in\mathbb R^{N\times 12}$ and $\gamma$ is the compression exponent. 
By emphasizing pitch-related cues such as notes and chords and remaining relatively insensitive to high-frequency noise, the chroma spectrum offers a musically informed representation well suited to music enhancement. 
The chroma loss is defined as the MSE between the chroma spectra of the clean and enhanced music, i.e., 
\begin{equation}
    \mathcal L_{Chroma} = \mathbb{E}_{(\bm{R},\bm{\hat{R}})}\lVert  \bm{R}- \bm{\hat{R}}\rVert_{F}^{2}.
\end{equation}
The chroma loss guides the model to preserve tonal structure, yielding more natural and perceptually pleasing enhanced music, where $\lVert \cdot \rVert_{F}$ is the Frobenius norm.

\item {}{\textbf{Phase Loss}:}
Since direct optimization of the phase spectrum is hampered by phase wrapping, inspired by \cite{ai2023neural}, we apply an anti-wrapping function
$f_{AW}(t)=\bigl|t-2\pi \operatorname{round}(\tfrac{t}{2\pi})\bigr|$
to the phase error to explicitly optimize the phase spectrum, and define the phase loss as
\begin{equation}
    \mathcal L_{P}=\sum_{\theta \in \Theta}^{} \mathbb{E}_{(\bm{P},\bm{\hat{P}})}\lVert f_{AW}(\theta( \bm{P}- \bm{\hat{P}}))\rVert_{1},
\end{equation}
where $\Theta$ denotes a three-element set of phase operations: the identity operation, group delay, and instantaneous angular frequency. 
$\lVert \cdot \rVert_{1}$ is the L1 norm.
\end{itemize}

\begin{table*}[t]
    \centering
    \caption{Objective evaluation results among DSME and baselines for different tasks.}
    \resizebox{0.9\textwidth}{!}{
    \begin{tabular}{cccccccc}
    \specialrule{1.2pt}{0pt}{0pt}
         Task & Model & fwSSNR$\uparrow$ & MRS$\downarrow$ & L1-SD$\downarrow$ & FAD$\downarrow$ & LSD$\downarrow$ & ViSQOL$\uparrow$ \\ \hline
         \multirow{5}*{Music Denoising}& Distorted Music (No Enhancement) & 4.37 & 2.52 & 3.98 & 8.95 & 1.84 & 4.15 \\
         & Mel2Mel+DiffWave \cite{kandpal2022music} & 6.84 & 1.86 & 2.67 & 1.39 & 1.26 & 4.25 \\
         & TFC-CPq \cite{chae2023exploiting}& 13.35 & 1.12 & 1.79 & 0.32 & 0.84 & 4.52 \\
         & MP-SENet \cite{lu2025explicit}& 14.22 & 1.13 & 1.77 & 0.42 & 0.81 & 4.53 \\
         & DSME (Proposed) & \textbf{14.80} & \textbf{1.04} & \textbf{1.70} & \textbf{0.30} & \textbf{0.78} & \textbf{4.55}  \\ \hline
         \multirow{5}*{Music Dereverberation}& Distorted Music (No Enhancement) & 9.97 & 1.23 & 1.52 & {0.39} & 0.81 & 4.49 \\
         & Mel2Mel+DiffWave \cite{kandpal2022music} & 7.77 & 1.77 & 2.45 & 1.15 & 1.13 & 4.30 \\
         & TFC-CPq \cite{chae2023exploiting}& 7.95 & 1.65 & 1.99 & 1.88 & 0.97 & 4.39 \\
         & MP-SENet \cite{lu2025explicit}& 11.58 & 1.17 & 1.65 & 0.57 & 0.77 & 4.50 \\
         & DSME (Proposed) & \textbf{12.15} & \textbf{1.04} & \textbf{1.44} & \textbf{0.30} & \textbf{0.68} & \textbf{4.53}  \\ \hline
         \multirow{5}*{Mixed-Distortion Music Enhancement}& Distorted Music (No Enhancement) & 2.18 & 2.95 & 4.16 & 10.11 & 1.91 & 4.07 \\
         & Mel2Mel+DiffWave \cite{kandpal2022music} & 5.97 & 2.00 & 2.77 & 1.72 & 1.33 & 4.22 \\
         & TFC-CPq \cite{chae2023exploiting}& 5.33 & 1.94 & 2.64 & 2.57 & 1.17 & 4.31 \\
         & MP-SENet \cite{lu2025explicit}& 8.50 & 1.72 & 2.36 & \textbf{1.27} & 1.03 & 4.39 \\
         & DSME (Proposed) & \textbf{8.84} & \textbf{1.67} & \textbf{2.28} & 1.39 & \textbf{1.02} & \textbf{4.40}  \\ 
         \specialrule{1.2pt}{0pt}{0pt}
    \end{tabular}}
    \label{tab:1}
\end{table*}

\vspace{-2mm}
\subsubsection{Training Process}
\vspace{-2mm}
The overall generator loss is formulated as a linear combination of the previously described generative loss and spectral-level loss, i.e.,
\begin{equation}
    \mathcal L=\mathcal L_{G}+\mathcal L_{FM} + \lambda_1\mathcal L_{A}+\lambda_2\mathcal L_{Chroma}+\lambda_3\mathcal L_{P},
\end{equation}
where $\lambda_1$, $\lambda_2$ and $\lambda_3$ are hyperparameters.
The training of the DSME follows the standard training process of GAN, i.e., using $\mathcal L$ and $\mathcal L_{D}$ to train the generator and discriminator alternately.

\vspace{-2mm}
\section{Experiments}
\vspace{-2mm}

\vspace{-2mm}
\subsection{Dataset}
\vspace{-2mm}
In our experiments\footnote{Audio samples are available at: \url{https://anonymity225.github.io/DSME/}.}, we used Medley-solos-DB dataset \cite{lostanlen2018medley} as the clean music dataset.
Following prior work \cite{kandpal2022music,chae2023exploiting}, we removed the low-quality distorted electric guitar category and split the rest into training, validation, and test sets, yielding 5,437, 2,999, and 11,281 utterances, respectively. 
We then created three types of distorted samples, defining three corresponding enhancement tasks.
All utterances were downsampled to 16 kHz (i.e., $f_s=16000$) for the experiments.
\begin{itemize}[leftmargin=*]
\item {}{\textbf{Music Denoising}:}
We constructed the noisy data by adding noise from the DEMAND dataset \cite{thiemann2013diverse}, sampling the signal-to-noise ratio (SNR) randomly between 0 and 5 dB to control the noise intensity.

\item {}{\textbf{Music Dereverberation}:}
We simulated reverberant conditions by convolving the clean data with room impulse responses from the DNS Challenge dataset \cite{dubey2024icassp}.

\item {}{\textbf{Mixed-Distortion Music Enhancement}:}
We constructed the mixed-distortion data by combining the above two corruption processes.
\end{itemize}

\begin{table}[t]
    \centering
    \caption{Average preference (\%) in ABX tests on mixed-distortion enhancement, where N/P stands for ``no preference" and $p$ denotes the $p$-value of a $t$-test between two models.}
    \resizebox{0.9\columnwidth}{!}{
    \begin{tabular}{cccccc}
    \specialrule{1.2pt}{0pt}{0pt}
         DSME & Mel2Mel+DiffWave & TFC-CPq & MP-SENet & N/P & p \\
         \hline
         \textbf{68.79} & 22.59 & - & - & 8.62 & \textbf{$\bm{<}$0.01} \\
         \textbf{71.38} & - & 17.76 & - & 10.86 & \textbf{$\bm{<}$0.01} \\
         \textbf{63.21} & - & - & 26.07 & 10.72 & \textbf{$\bm{<}$0.01} \\ 
         \specialrule{1.2pt}{0pt}{0pt}
    \end{tabular} }
    \label{tab:2}
\end{table}

\vspace{-2mm}
\subsection{Model Details}
\vspace{-2mm}
For the DSME generator, we computed STFT with a 25 ms window, 6.25 ms hop, and a 400-point FFT (i.e., $K=201$); all 2D convolutions used $1\times 3$ kernels.
For the discriminator, we used CQT with 7 octaves (i.e., $M=7$) and 12 bins per octave (i.e., $B=12$), with $f_{min}=55$Hz; transposed and standard 2D convolutions used $3\times 9$ and $4\times 4$ kernels, respectively. Loss weights $\lambda_1$, $\lambda_2$ and $\lambda_3$ were set to 0.9, 0.2 and 0.3, with $\gamma=0.5$.
We trained DSME for 200 epochs using AdamW on a single NVIDIA RTX A800 GPU ($\beta_{1}=0.8,\beta_{2}=0.99$, weight decay 0.01). The learning rate started at 0.0005 and decayed by 0.99 each epoch.

We compared DSME with Mel2Mel+DiffWave \cite{kandpal2022music}, TFC-CPq \cite{chae2023exploiting} and the speech enhancement model MP-SENet \cite{lu2025explicit}.
For fairness, all baselines were retrained on our music dataset using official implementations.

\vspace{-2mm}
\subsection{Evaluation Metrics}
\vspace{-2mm}
To comprehensively assess music enhancement performance, we employed four widely used objective metrics \cite{hu2007evaluation,roblek2019fr}: frequency-weighted segmental SNR (fwSSNR), multi-resolution spectral loss (MRS), L1 spectral distance (L1-SD), and Fréchet audio distance (FAD). 
In addition, we included two objective measures commonly used in speech enhancement, log spectral distance (LSD) and virtual speech quality objective listener (ViSQOL) \cite{chinen2020visqol}, as auxiliary metrics. 

We also ran ABX preference tests on Amazon Mechanical Turk for the mixed-distortion task, comparing DSME pairwise with each baseline. For each comparison, 20 enhanced test utterances were randomly selected and rated by at least 30 native English-speaking listeners, who chose the better sample or no preference (N/P). We report average preference scores and $t$-test $p$-values for significance.

\vspace{-2mm}
\subsection{Experimental Results and Analysis}
\vspace{-2mm}

As shown in Table \ref{tab:1}, DSME consistently delivered the best or second-best performance across all objective metrics and tasks.
For the music denoising task, DSME achieved the highest fwSSNR and ViSQOL scores, along with the lowest MRS, L1-SD, FAD, and LSD scores, demonstrating its strong noise suppression capability.
Despite strong STFT-domain baselines (TFC-CPq and MP-SENet), DSME still outperforms them overall, highlighting the effectiveness of our CQT-based discriminator and chroma-aware design for music enhancement.

\begin{table}[t]
    \centering
    \caption{Objective evaluation results among DSME and its ablated variants on mixed-distortion music enhancement task.}
    \resizebox{0.9\columnwidth}{!}{
    \begin{tabular}{ccccc}
    \specialrule{1.2pt}{0pt}{0pt}
         Model & fwSSNR$\uparrow$ & MRS$\downarrow$ & L1-SD$\downarrow$ & FAD$\downarrow$ \\ \hline
         DSME & \textbf{8.84} & \textbf{1.67} & 2.28 & 1.39 \\
         DSME w/o CQT & 8.61 & 1.72 & 2.32 & 1.33 \\
         DSME w/o Chroma & 8.49 & 1.69 & \textbf{2.26} & \textbf{1.30} \\ 
         \specialrule{1.2pt}{0pt}{0pt}
    \end{tabular}}
    \label{tab:3}
\end{table}

In the music dereverberation task, DSME again surpassed all baselines on all objective metrics, as shown in Table \ref{tab:1}. 
Interestingly, the reverberant music (i.e., no enhancement) and DSME-enhanced music had comparable FAD values, which is likely because reverberation is less destructive to the global spectral distribution than additive noise. 
In contrast, the other baselines yielded much higher FAD scores, indicating that they not only failed to effectively remove reverberation but also introduced additional spectral artifacts.

For the mixed-distortion music enhancement task, where both noise and reverberation are present, DSME achieved clear gains over all baselines on most objective metrics, although its FAD was slightly worse than that of MP-SENet. 
To provide additional perceptual evidence, we therefore conducted ABX preference listening tests. 
The subjective results are listed in Table \ref{tab:2} and show that listeners significantly preferred DSME over each baseline model ($p<0.01$), indicating consistently higher perceived music quality. 
Overall, these objective and subjective results confirm that DSME’s strategy of using dual time–frequency spectral representations for prediction and discrimination in music enhancement is effective and particularly well aligned with the characteristics of music.

\vspace{-2mm}
\subsection{Ablation Studies}
\vspace{-2mm}


Relative to baselines, DSME targets music-specific characteristics via CQT-based discrimination and a chroma loss. To validate their contributions, we conduct ablations on the mixed-distortion task and report four music-oriented objective metrics.

Table \ref{tab:3} summarizes the results. Replacing the discriminator’s CQT input with STFT (DSME w/o CQT) degrades fwSSNR, MRS, and L1-SD, indicating that CQT-based discrimination provides more music-specific guidance and improves time-frequency structure reconstruction. Removing the chroma loss (DSME w/o Chroma) also degrades fwSSNR and MRS, confirming the benefit of chroma supervision. Although DSME w/o Chroma slightly lowers L1-SD, this likely reflects a trade-off with chroma regularization rather than better perceptual quality. The ablated variants also yield slightly lower FAD, possibly because FAD can favor more conservative outputs. 

\vspace{-1mm}
\section{Conclusion}
\vspace{-1mm}

We propose DSME, a neural music enhancement model that leverages dual time–frequency spectral representations tailored to music. DSME uses a GAN with STFT-based amplitude–phase spectra in the generator and an octave-segmented CQT spectrum in the discriminator. STFT’s fixed window and invertibility make prediction well-conditioned, while CQT’s log-frequency scale and variable window better capture octave-related harmonics. We further introduce a chroma-spectrum loss to preserve pitch and harmonic consistency. Experiments show DSME consistently outperforms strong baselines across multiple distortion scenarios in objective and subjective evaluations, validating the dual-spectrum design. Future work will extend DSME to more realistic in-the-wild recordings.

\bibliographystyle{IEEEtran}

\bibliography{mybib}


\end{document}